\documentclass{llncs}
\usepackage{booktabs} 
\usepackage{comment}
\usepackage{graphicx}
\usepackage[lofdepth,lotdepth]{subfig}
\usepackage{cellspace}
\usepackage{wrapfig,picins}
\usepackage[export]{adjustbox}

\usepackage[sorting=none, maxnames=50]{biblatex}
\usepackage{amsmath}
\usepackage{color}
\usepackage{url}
\usepackage{booktabs} 
\usepackage{amsfonts}
\DeclareGraphicsExtensions{.pdf,.png,.jpg}
\usepackage[english]{babel}
\usepackage[table,xcdraw]{xcolor}
\usepackage{multirow}

\usepackage[section]{placeins}

\usepackage{listings}
\usepackage{xcolor}
\definecolor{codegreen}{rgb}{0,0.6,0}
\definecolor{codegray}{rgb}{0.5,0.5,0.5}
\definecolor{codepurple}{rgb}{0.58,0,0.82}
\definecolor{backcolour}{rgb}{0.95,0.95,0.92}
\lstdefinestyle{mystyle}{
    backgroundcolor=\color{backcolour},   
    commentstyle=\color{codegreen},
    keywordstyle=\color{magenta},
    numberstyle=\tiny\color{codegray},
    stringstyle=\color{codepurple},
    basicstyle=\ttfamily\footnotesize,
    breakatwhitespace=false,         
    breaklines=true,                 
    captionpos=b,                    
    keepspaces=true,                 
    numbers=left,                    
    numbersep=5pt,                  
    showspaces=false,                
    showstringspaces=false,
    showtabs=false,                  
    tabsize=2
}
\begin{document}

\frontmatter          
\pagestyle{empty}  

\title{An example of use of Variational Methods in Quantum Machine Learning}
 \author{Marco Simonetti\inst{1} $^{ORCID: 0000-0003-2923-5519}$
 \newline Damiano Perri\inst{1} $^{ORCID: 0000-0001-6815-6659}$
 \newline Osvaldo Gervasi\inst{2} $^{ORCID: 0000-0003-4327-520X}$ 
 }
\institute{
University of Florence, Dept. of Mathematics and Computer Science, Florence, Italy \and University of Perugia, Dept. of Mathematics and Computer Science, Perugia, Italy
}
\titlerunning{An example of use of Variational Methods in Quantum Machine Learning} 
\authorrunning{Damiano Perri, Marco  Simonetti and Osvaldo Gervasi} 

\maketitle
\begin{abstract}
This paper introduces a deep learning system based on a quantum neural network for the binary classification of points of a specific geometric pattern (Two-Moons Classification problem) on a plane.

We believe that the use of hybrid deep learning systems (classical + quantum) can reasonably bring benefits, not only in terms of computational acceleration but in understanding the underlying phenomena and mechanisms; that will lead to the creation of new forms of machine learning, as well as to a strong development in the world of quantum computation.

The chosen dataset is based on a 2D binary classification generator, which helps test the effectiveness of specific algorithms; it is a set of 2D points forming two interspersed semicircles. It displays two disjointed data sets in a two-dimensional representation space: the features are, therefore, the individual points' two coordinates, $x_1$ and $x_2$.

The intention was to produce a quantum deep neural network with the minimum number of trainable parameters capable of correctly recognising and classifying points.

\end{abstract}


\section{Introduction}
The stages of knowledge in the history of humankind have always alternated between amazement at the immensity of the phenomenon before us and the joyful conquest for the objectives set.
The development and introduction into the science of a powerful mathematical arsenal have slowly enabled many obstacles to be overcome, confirming Galileo's intuition that mathematics is the straightforward language that enables us to dialogue with nature.
At the beginning of the 20th century, our optimism in positivist determinism was strongly shaken by new phenomena that led to the birth of quantum mechanics. Nevertheless, the observation of the generation of deterministic chaos from models with a seemingly simple apparatus of differential equations and the evident need to describe basic molecular structures by simulating them with automatic calculation tools that require ever-increasing computational capabilities are suggesting that the time has come for a profound reflection on our tools of scientific investigation.
Feynman stated that describing the reality that surrounds us, which has an intrinsically quantum nature, through a type of computation based on quantum mechanics would be the key to exponentially lowering the computational complexity of the system in question and correctly managing the predictive capacity for the model.
We might also add that the introduction of quantum computing machines would also solve the problems related to the imminent reaching of the construction limit of current computers (Amdahl's law), to the possibility of a drastic reduction in the energy required by today's computing machines thanks to computational reversibility, and to the development of nanotechnology.

The possibility of such opportunities has resulted in a growing international interest in such devices, with considerable funding in the relevant research areas.
Quantum Computing machines can therefore respond to questions that would be unimaginable for actual old-style apparatuses \cite{deutsch1992rapid}, as they would need the whole Universe's existence time to complete the same task. That is allowed by some quantum peculiarities that show up at the littlest of scales, like Superposition, Entanglement and Interference \cite{grover1997quantum}.
Although the theoretical study of quantum algorithms suggests the real possibility of solving computational problems that are currently intractable \cite{shor1999polynomial}, the current technology has not reached maturity, such as to obtain truly appreciable advantages. We are still in the so-called era of Noisy Intermediate-Scale Quantum (NISQ) \cite{preskill2018quantum}.
NISQ-devices are noisy and have limited quantum resources; this presents various obstacles when implementing a gate-based quantum algorithm. In order to determine if an implementation of a gate-based algorithm would run effectively on a particular NISQ-device, numerous aspects of circuit implementation, such as depth, width, and noise, should be considered \cite{leymann2020bitter, salm2020criterion}.

Variational Methods \cite{zeidler2013nonlinear, smith1998variational} are widely used in physics, and most of all in quantum mechanics \cite{borghi2018variational}. Their direct successors, Variational Quantum Algorithms (VQAs), have appeared to be the most effective technique for gaining a quantum advantage on NISQ devices. VQAs are undoubtedly the quantum equivalent of very effective machine-learning techniques like neural networks.
Furthermore, VQAs use the classical optimisation toolbox since they employ parametrised quantum circuits to run on the quantum computer and then outsource parameter optimisation to a classical optimiser. In contrast to quantum algorithms built for the fault-tolerant time period, this technique provides the extra benefits of keeping the quantum circuit depth small and minimising noise \cite{cerezo2021variational}.

\begin{figure}[h]
    \centering
    \includegraphics[width=0.5\textwidth]{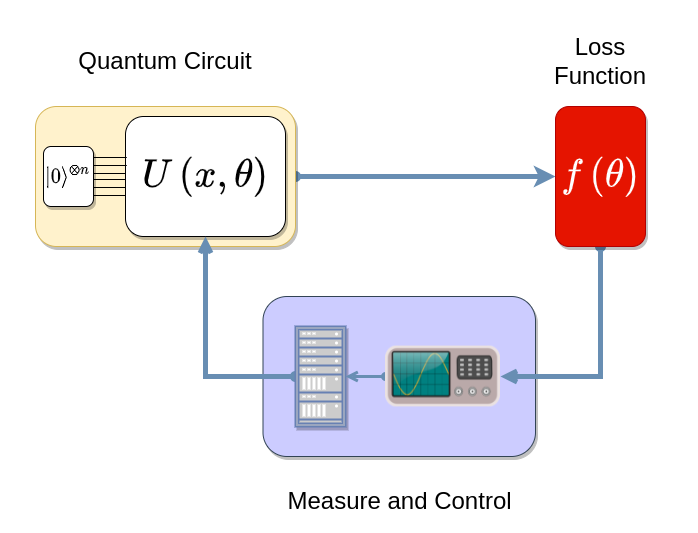}
    \caption{Scheme}
    \label{fig:img01}
\end{figure}

The aim is to find the exact, or a well-approximated, parameter values' set that minimises a given cost (loss) function, which depends on the parameters themselves and, naturally, on the input values, \textit{x}; these are the non-trainable part of the schema. An output measurement apparatus and a parameter modulation circuit serve as regulators (Fig.\ref{fig:img01}).
In our specific case, we are dealing with a \textit{n-qubits} quantum ansatz, whose transformation is represented by the $n\times n$ unitary matrix \textit{U} that depends on the vector of parameters $\theta$, with dimensions $m\;\epsilon\;\mathbb{N}$ (eq. \eqref{eq1}); the loss function we refer to is the \textit{expectation value} measured on every single qubit channel (eq. \eqref{eq3}); the observable is the operator $O_i$, where $Z_i$ is the Z-Pauli Operator, acting on i-th qubit (eq. \eqref{eq2}); the quantum state on which to operate the measurements will be the state resulting from the ansatz ($\psi$ in eq. \eqref{eq1}).
\begin{equation}\label{eq1}
    \Big|\psi\left(\textbf{$\overline{\theta}$}\;\right)\Big\rangle=U\left(\textbf{$\overline{\theta}$}\;\right)\cdot|0\rangle^{\otimes n}
\end{equation}
\begin{equation}\label{eq2}
    O_i = Z_i\:, \hspace{1em}i\;\epsilon\:\left\{1..n  \right\}\subset \mathbb{N}
\end{equation}
\begin{equation}\label{eq3}
    \Big\langle E_i\left(\textbf{$\overline{\theta}$}\;\right)\Big\rangle = \left\langle \psi^\dagger\left(\textbf{$\overline{\theta}$}\;\right)\, \Big|\;O_i\,\Big|\;\psi\left(\textbf{$\overline{\theta}$}\;\right)\right\rangle 
\end{equation}
It is well-known that Quantum Computers can take care of specific issues quicker than traditional ones. As it may, stacking information into a quantum computer is not paltry. It should be encoded in quantum bits (qubits) to stack the information. There are multiple ways qubits can address the data, and, in this manner, various information encodings are conceivable \cite{weigold2020data}. Several methods can embed data: Basis Encoding \cite{vedral1996quantum, cortese2018loading}, Amplitude Encoding \cite{larose2020robust, harrow2009quantum, schuld2017implementing}, and Angle Encoding \cite{weigold2021expanding, grant2018hierarchical} are the most common to implant information into a firstly prepared quantum state.

Our work aims to explore aspects of the usability of quantum computation in machine learning. Specifically, we sought to compare the performance of a basic FFNN with dense layers and an analogous network composed of quantum components in learning a simple classification problem.

\section{Related works}
The possibility of using machine learning techniques in quantum computing has been gaining ground since 2010 \cite{lloyd2013quantum, wiebe2014quantum, lloyd2014quantum}.
The incorporation of quantum algorithms into machine learning programmes is known as quantum machine learning \cite{perri2020binary, benedetti2020skin, perri2021new, perri2021synthetic}. The expression is most commonly used to refer to machine learning methods for evaluating classical data run on a quantum computer. While machine learning techniques are used to calculate vast quantities of data, quantum machine learning employs qubits and quantum operations or specialised quantum systems to speed up computing \cite{yoo2014quantum} and data storage \cite{ouyang2021quantum}.
Quantum machine learning also refers to an area of study investigating the theoretical and functional analogies between specific physical systems and learning systems, namely neural networks \cite{wan2017quantum}.

Unfortunately, the current technological development does not yet allow the full potential of quantum computers to be expressed, which will only reach maturity in the next few decades. At present, however, it is possible to use quantum computers in feedback circuits that mitigate the effect of various noise components \cite{bharti2022noisy, endo2021hybrid}. VQAs, which utilise a conventional optimiser to train a parameterised quantum circuit, have been considered an effective technique for addressing these restrictions.

The core part of the computational stage consists of a sequence of gates applied to specified wires in variational circuits. Like the design of a neural network that merely describes the basic structure, the variational approach can optimise the types of gates and their free parameters.
In time, the quantum computing community has suggested several variational circuit types that can distinguish three main base structures, depending on the shape of the ansatz: layered gate ansatz \cite{schuld2019quantum}, alternating operator ansatz \cite{farhi2014quantum, verdon2017quantum}, and tensor network ansatz \cite{huggins2019towards}.

The first type of architecture inspired us in implementing our network, trying to keep the network simple, with a minimum number of quantum gates and trainable parameters.

\section{The architecture of the system}
\begin{figure}[h]
    \centering
    \includegraphics[width=1.0\textwidth]{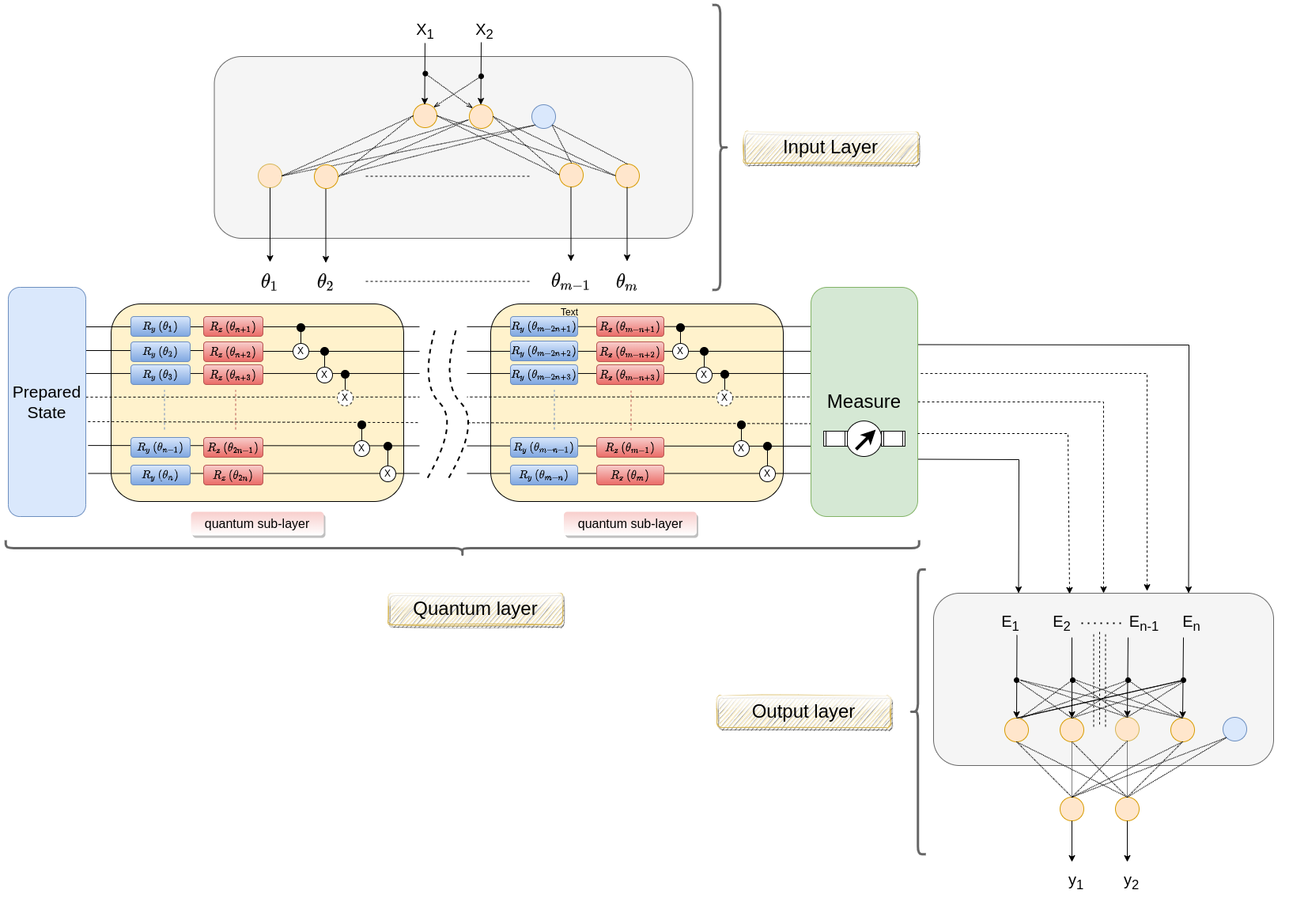}
    \caption{Architecture of the quantum neural network}
    \label{fig:img02}
\end{figure}

Our network has a general structure of this type:
\begin{itemize}
    \item A first classical dense layer, to accept features as input
    \item A quantum circuit formed by a succession of rotational and entanglement gates
    \item A final classical dense layer for classification
\end{itemize}
Our quantum structure differs from some other proposals, which encodes input features to transform them into quantum states \cite{romero2017quantum, schuld2020circuit}; our initial state, the "prepared state", is simply $\vert 0\rangle^{\otimes n}$. Our architecture is based on the typical "Prepared State fixed; Ansatz parametrised" model. Several variants exist in the literature, especially the most known \textit{Instantaneous Quantum Polynomial} (IQP) circuits \cite{shepherd2008instantaneous}. The initial dense network, which we called \textit{input-layer}, acts as an interface between the input data and the quantum layer; its output directly drives the internal rotations of the quantum state, attributing new values to the rotation angles at each epoch (Fig. \ref{fig:img02}).
Almost all the simulations were conducted using quantum sub-layers consisting of a first stack of rotation gates around the Y-axis of the Bloch's sphere, a second stack of rotation gate around the Z-axis and a chain of \textit{cX} gates (CNOT) to maximise the entanglement effect. Rotation operators help primarily to ensure the \textit{expressibility} of a parameterized quantum circuit. That is essentially the total coverage of Hilbert space by the hypothesis space of the ansatz itself \cite{sim2019expressibility}.
In addition, the chain of CNOT operators is tasked with maximising the entanglement effect of individual qubits, as the entanglement phenomenon plays a crucial role in quantum computation. This feature is called \textit{Entangling Capability} \cite{hubregtsen2021evaluation}.

An authentic advantage is that the network's response remains independent of the number of qubits. The only real quantum noise remains due to the construction and constitution of the ansatz and the device, or simulator, used for experimentation.

\subsection{Data extraction and processing}
The database chosen for the classification problem is called \textit{two-moons}, which generates two interleaving semicircles of 2D-points, and is characteristic for the study of clustering and classification algorithms.

The dataset was generated thanks to the dedicated function from the Python Scikit-Learn library\footnote{https://scikit-learn.org/stable/index.html}.

\begin{figure}[h]
    \centering
    \includegraphics[width=1.0\textwidth]{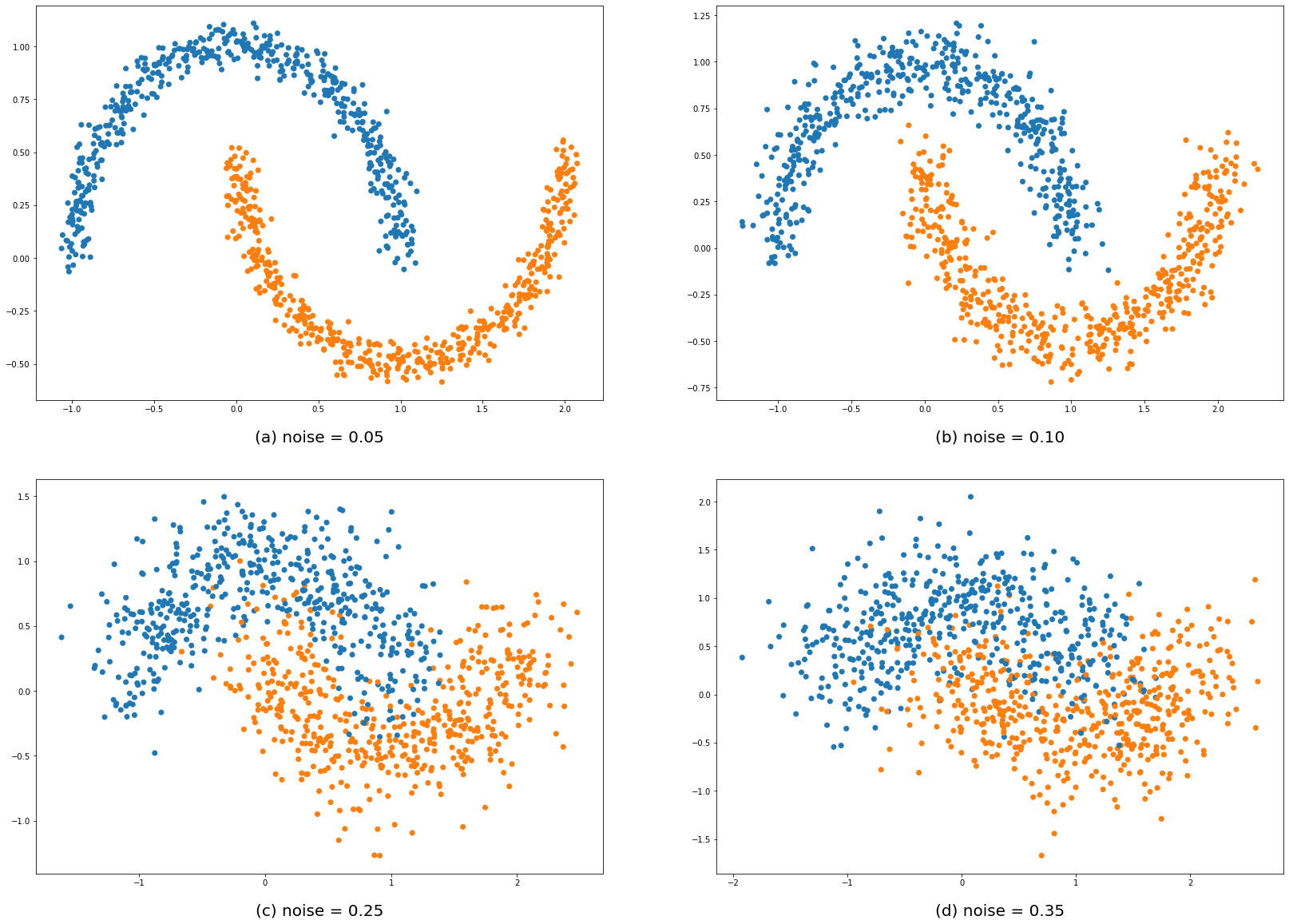}
    \caption{Distribution of pattern points on the plane in relation to Gaussian Noise}
    \label{fig:img03}
\end{figure}

The input domain is $\left[-2.0, 3.0\right] \times \left[-2.0, 2.5\right] \subset \mathbb{R}^2$.
One thousand samples were randomly generated, equally distributed over the two classes. The classes were initially chosen with a Gaussian Error Coefficient of 0.05: this coefficient quantifies the capacity of separation of the two categories: the higher its value, the greater the degree of overlap of the two classes (Fig. \ref{fig:img03}).
The number of samples in the dataset was divided in such a way as to ensure the following quotas: 70\% for the training set, 20\% for the validation set and the remaining 10\% for the test set.

\subsection{Model construction and validation}
Two twin models have been realised in parallel, using for one of them the Python library for quantum simulation Cirq\footnote{https://quantumai.google/cirq} with TensorFlowQuantum\footnote{https://www.tensorflow.org/quantum}, by Google, and for the other an open-source Python library, PennyLane\footnote{https://pennylane.ai/}, by Xanadu, which can be perfectly interfaced with the most famous Deep Learning frameworks, such as Keras, and of course TensorFlow, and PyTorch.

All tests and simulations were performed using the specialised open-source library for Deep Learning, Keras\footnote{https://keras.io/}.

\begin{figure}[h!]
    \centering
    \subfloat[TQF Model]{
        \includegraphics[width=0.6\linewidth]{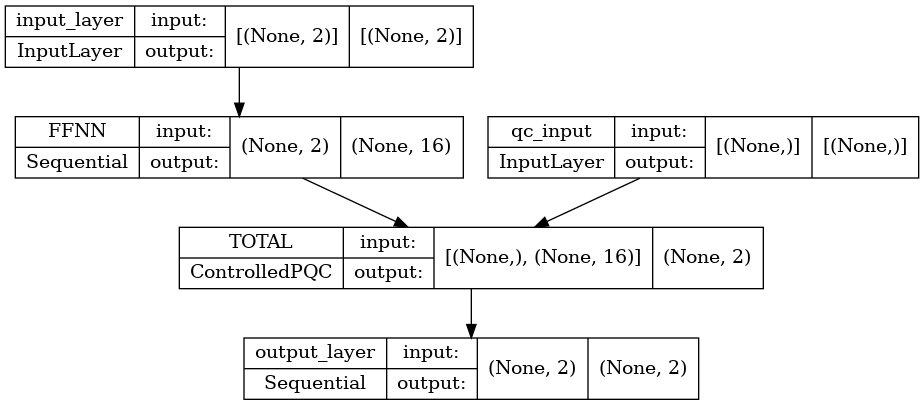}
        \label{fig:img04a}}
    \hfill
    \subfloat[PQML Model]{
        \includegraphics[width=0.3\linewidth]{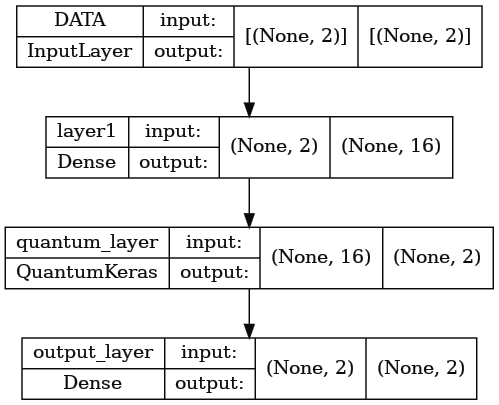}
        \label{fig:img04b}}
    \caption{Quantum Neural Network Models utilised}
    \label{figure:img04}
\end{figure}

The TFQ (TensorFlowQuantum) model is composed as follows (Fig. \ref{fig:img04a}):
\begin{itemize}
    \item A layer for data input, called \textit{input\_layer}.
    \item A Dense layer, which adapts the input to the number of parameters, in our specific case 16, called \textit{FFNN}.
    \item A layer for the ansatz input in the form of a tensor, called \textit{qc\_layer}.
    \item A control layer (called \textit{TOTAL}), which receives the two inputs, modulates the values of the parameters and returns, for each wire in the circuit, the expected value.
    \item A final Dense layer (called \textit{output\_layer}), with a number of outputs equal to the number of classes in the dataset, which returns the probability of belonging to them (\textit{softmax} activation).
\end{itemize}
The PQML (PennyLane Quantum Machine Learning) model is composed as follows (Fig. \ref{fig:img04b}):
\begin{itemize}
    \item A layer for data input, called \textit{DATA}.
    \item A Dense layer, which adapts the input to the number of parameters, in our specific case 16, called \textit{layer1}.
    
    \item A custom layer, called \textit{quantum\_layer}, designed as a subclass of the Layer class from the Keras library, which receives in input the values of the parameters of the ansatz and returns, for each wire of the circuit, the expected value.
    \item A final Dense layer (called \textit{output\_layer}), with a number of outputs equal to the number of classes in the dataset, which returns the probability of belonging to them (\textit{softmax} activation).
\end{itemize}

The number of ansatz parameters is given by the product of the number of qubits, the number of rotation operators acting on each line of the circuit within each sublayer, and the number of sublayers.
For our tests and simulations, we have opted to use an ansatz consisting of 4 sublayers to avoid creating an excessive computational load on the quantum layer.
By initially constraining the number of qubits to 2 and not introducing any form of noise, we can say that both models proved to be extremely light in terms of the number of trainable parameters: 54 against the 354 of a similar FFNN Fully-Connected, with three layers.
The disadvantage is that the simulation run times are about one order of magnitude longer than the latter. In fact, to simulate an ansatz is necessary to operate many matrix products against only one matrix product for a standard dense layer.

The entire model was compiled using a classical optimiser, \textit{SGD} (Stochastic Gradient Descent), with \textit{Mean Absolute Error} as the Loss Function and \textit{Accuracy} as the metric.

\subsection{Analysis of results}
\begin{figure}[h]
    \centering
    \includegraphics[width=1.0\textwidth]{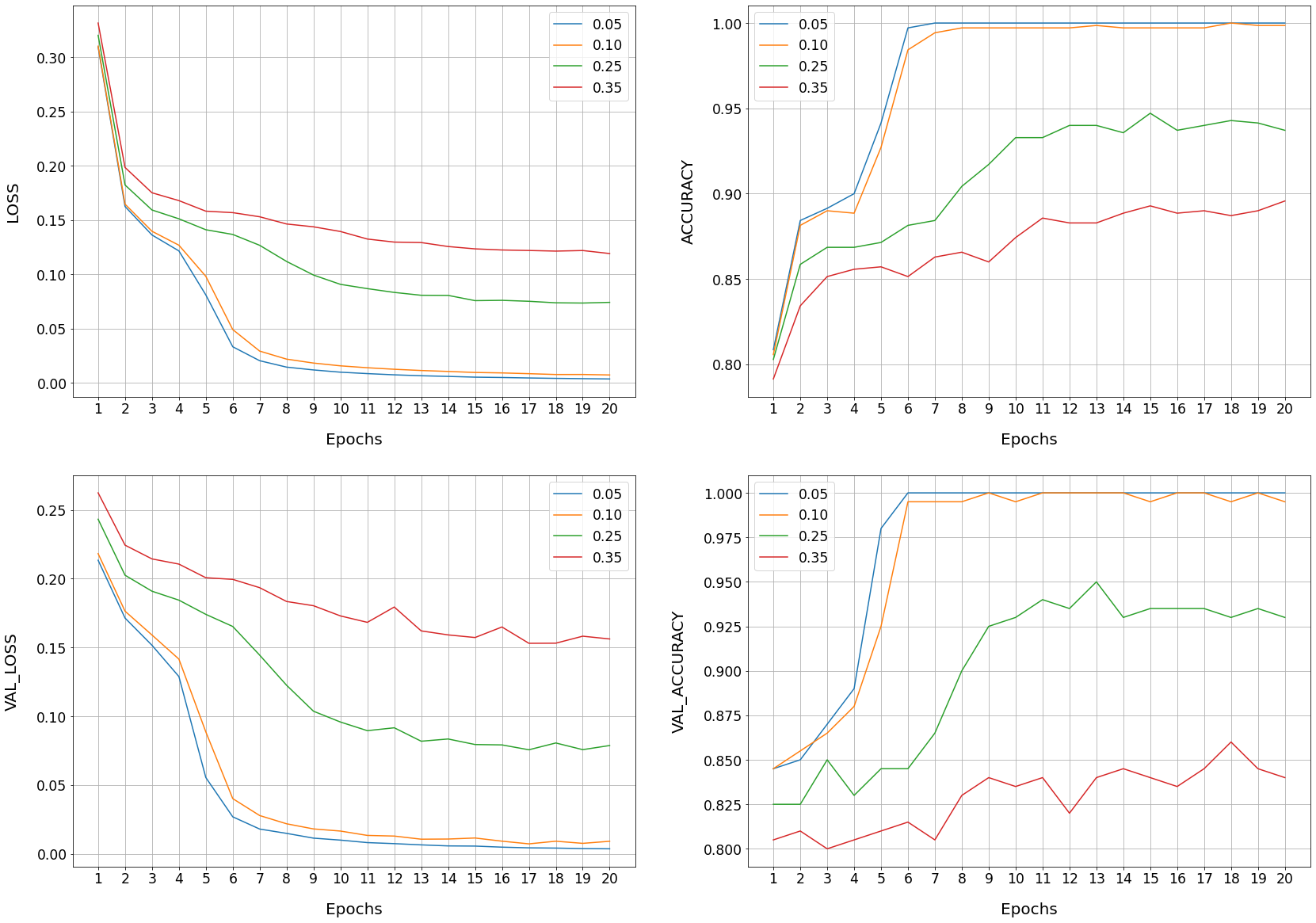}
    \caption{Metrics for TFQ model}
    \label{fig:img05}
\end{figure}

The results obtained are excellent. Figure \ref{fig:img05} shows the metrics (LOSS: loss function on training set, ACCURACY: accuracy on training set, VAL\_LOSS: loss function on validation set, VAL\_ACCURACY: accuracy on validation set,) for the TFQ model, trained for 20 epochs, with different values of Gaussian Noise on the dataset generation (legend).

The model achieves 100\% accuracy for low-noise datasets and more than 80\% accuracy for high-noise datasets.

\begin{table}[]
\centering
\begin{tabular}{c|rrr|rrr|rrr|}
\cline{2-10}
\multicolumn{1}{l|}{} &
  \multicolumn{3}{c|}{\cellcolor[HTML]{656565}{\color[HTML]{FFFE65} \textbf{Samples: 20 / 200}}} &
  \multicolumn{3}{c|}{\cellcolor[HTML]{656565}{\color[HTML]{FFFE65} \textbf{Samples: 100 / 1,000}}} &
  \multicolumn{3}{c|}{\cellcolor[HTML]{656565}{\color[HTML]{FFFE65} \textbf{Samples: 500 / 5,000}}} \\ \hline
\rowcolor[HTML]{C0C0C0} 
\multicolumn{1}{|c|}{\cellcolor[HTML]{A5E4F7}} &
  \multicolumn{3}{c|}{\cellcolor[HTML]{C0C0C0}\textbf{n qubits}} &
  \multicolumn{3}{c|}{\cellcolor[HTML]{C0C0C0}\textbf{n qubits}} &
  \multicolumn{3}{c|}{\cellcolor[HTML]{C0C0C0}\textbf{n qubits}} \\ \cline{2-10} 
\rowcolor[HTML]{C0C0C0} 
\multicolumn{1}{|c|}{\multirow{-2}{*}{\cellcolor[HTML]{A5E4F7}\textbf{NOISE}}} &
  \multicolumn{1}{c|}{\cellcolor[HTML]{C0C0C0}\textbf{2}} &
  \multicolumn{1}{c|}{\cellcolor[HTML]{C0C0C0}\textbf{3}} &
  \multicolumn{1}{c|}{\cellcolor[HTML]{C0C0C0}\textbf{4}} &
  \multicolumn{1}{c|}{\cellcolor[HTML]{C0C0C0}\textbf{2}} &
  \multicolumn{1}{c|}{\cellcolor[HTML]{C0C0C0}\textbf{3}} &
  \multicolumn{1}{c|}{\cellcolor[HTML]{C0C0C0}\textbf{4}} &
  \multicolumn{1}{c|}{\cellcolor[HTML]{C0C0C0}\textbf{2}} &
  \multicolumn{1}{c|}{\cellcolor[HTML]{C0C0C0}\textbf{3}} &
  \multicolumn{1}{c|}{\cellcolor[HTML]{C0C0C0}\textbf{4}} \\ \hline
\multicolumn{1}{|c|}{\cellcolor[HTML]{A5E4F7}\textbf{0.05}} &
  \multicolumn{1}{r|}{100.0 \%} &
  \multicolumn{1}{r|}{95.0 \%} &
  95.0 \% &
  \multicolumn{1}{r|}{100.0 \%} &
  \multicolumn{1}{r|}{100.0 \%} &
  100.0 \% &
  \multicolumn{1}{r|}{100.0 \%} &
  \multicolumn{1}{r|}{100.0 \%} &
  100.0 \% \\ \hline
\multicolumn{1}{|c|}{\cellcolor[HTML]{A5E4F7}\textbf{0.10}} &
  \multicolumn{1}{r|}{100.0 \%} &
  \multicolumn{1}{r|}{95.0 \%} &
  95.0 \% &
  \multicolumn{1}{r|}{100.0 \%} &
  \multicolumn{1}{r|}{100.0 \%} &
  100.0 \% &
  \multicolumn{1}{r|}{100.0 \%} &
  \multicolumn{1}{r|}{100.0 \%} &
  100.0 \% \\ \hline
\multicolumn{1}{|c|}{\cellcolor[HTML]{A5E4F7}\textbf{0.25}} &
  \multicolumn{1}{r|}{95.0 \%} &
  \multicolumn{1}{r|}{80.0 \%} &
  90.0 \% &
  \multicolumn{1}{r|}{93.0\%} &
  \multicolumn{1}{r|}{94.0 \%} &
  94.0 \% &
  \multicolumn{1}{r|}{93.8 \%} &
  \multicolumn{1}{r|}{93.8 \%} &
  90.0 \% \\ \hline
\multicolumn{1}{|c|}{\cellcolor[HTML]{A5E4F7}\textbf{0.35}} &
  \multicolumn{1}{r|}{80.0 \%} &
  \multicolumn{1}{r|}{75.0 \%} &
  85.0 \% &
  \multicolumn{1}{r|}{88.0\%} &
  \multicolumn{1}{r|}{86.0 \%} &
  89.0 \% &
  \multicolumn{1}{r|}{88.4 \%} &
  \multicolumn{1}{r|}{88.6 \%} &
  88.4 \% \\ \hline
\end{tabular}
\caption{Percentage of correct evaluations on the test set for TFQ model.}
\label{tbl:tab01}
\end{table}

Table \ref{tbl:tab01} shows the percentages of correct evaluations on the test set as a function of the Gaussian Noise on the dataset, the number of qubits used, datasets with different cardinalities (the header of the table shows the number of samples of the test set on the number of total samples). It is referred to TFQ model.

\begin{figure}[h!]
    \centering
    \includegraphics[width=1.0\textwidth]{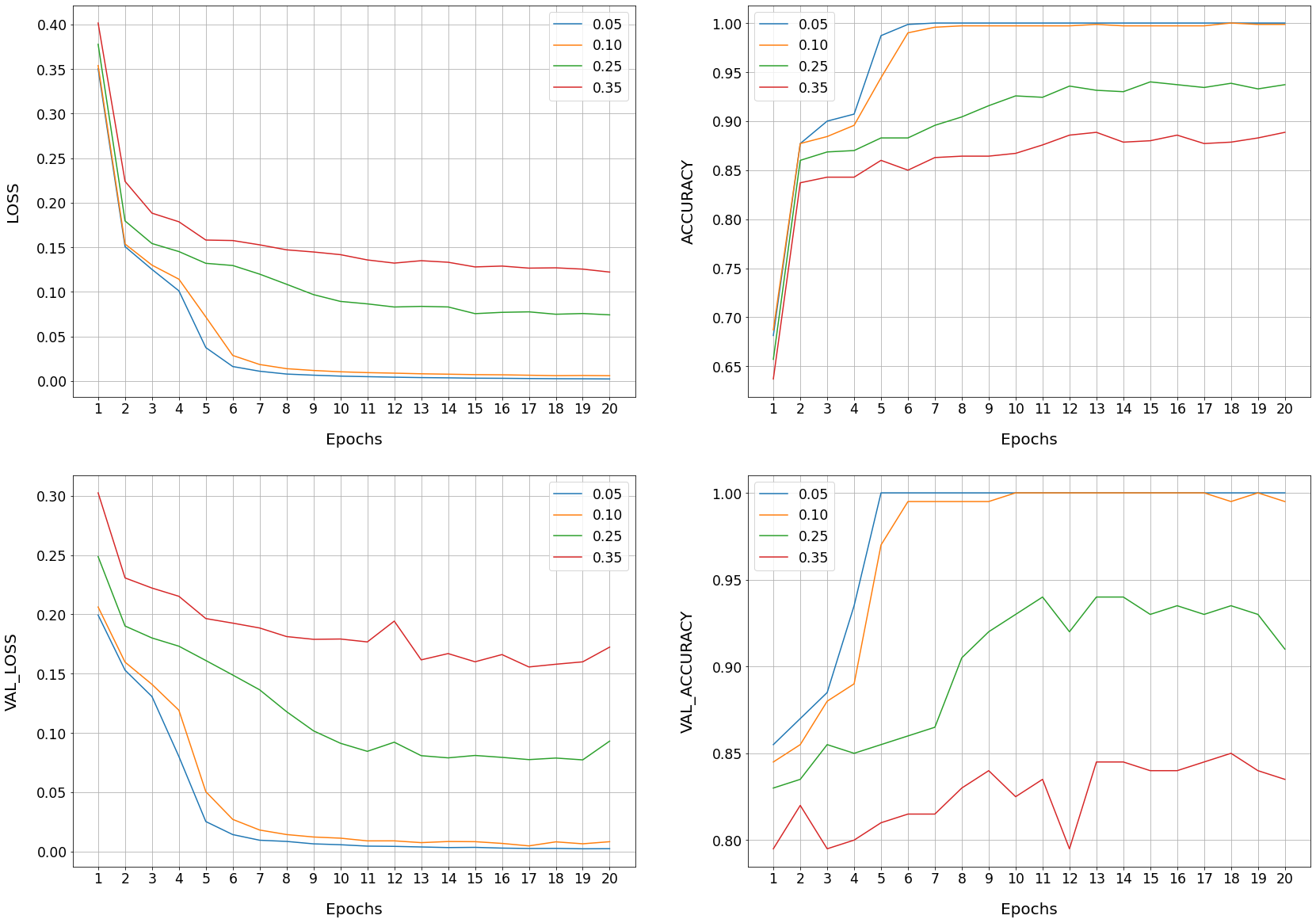}
    \caption{Metrics for PQML model}
    \label{fig:img06}
\end{figure}

Figure \ref{fig:img06} shows the usual metrics for the PQML model, trained for 20 epochs, with different values of Gaussian Noise on the dataset generation (legend).

The model achieves 100\% accuracy for low-noise datasets and beyond 80\% accuracy for high-noise datasets.

\begin{table}[]
\centering
\begin{tabular}{c|rrr|rrr|rrr|}
\cline{2-10}
\multicolumn{1}{l|}{} &
  \multicolumn{3}{c|}{\cellcolor[HTML]{656565}{\color[HTML]{FFFE65} \textbf{Samples: 20 / 200}}} &
  \multicolumn{3}{c|}{\cellcolor[HTML]{656565}{\color[HTML]{FFFE65} \textbf{Samples: 100 / 1,000}}} &
  \multicolumn{3}{c|}{\cellcolor[HTML]{656565}{\color[HTML]{FFFE65} \textbf{Samples: 500 / 5,000}}} \\ \hline
\rowcolor[HTML]{C0C0C0} 
\multicolumn{1}{|c|}{\cellcolor[HTML]{A5E4F7}} &
  \multicolumn{3}{c|}{\cellcolor[HTML]{C0C0C0}\textbf{n qubits}} &
  \multicolumn{3}{c|}{\cellcolor[HTML]{C0C0C0}\textbf{n qubits}} &
  \multicolumn{3}{c|}{\cellcolor[HTML]{C0C0C0}\textbf{n qubits}} \\ \cline{2-10} 
\rowcolor[HTML]{C0C0C0} 
\multicolumn{1}{|c|}{\multirow{-2}{*}{\cellcolor[HTML]{A5E4F7}\textbf{NOISE}}} &
  \multicolumn{1}{c|}{\cellcolor[HTML]{C0C0C0}\textbf{2}} &
  \multicolumn{1}{c|}{\cellcolor[HTML]{C0C0C0}\textbf{3}} &
  \multicolumn{1}{c|}{\cellcolor[HTML]{C0C0C0}\textbf{4}} &
  \multicolumn{1}{c|}{\cellcolor[HTML]{C0C0C0}\textbf{2}} &
  \multicolumn{1}{c|}{\cellcolor[HTML]{C0C0C0}\textbf{3}} &
  \multicolumn{1}{c|}{\cellcolor[HTML]{C0C0C0}\textbf{4}} &
  \multicolumn{1}{c|}{\cellcolor[HTML]{C0C0C0}\textbf{2}} &
  \multicolumn{1}{c|}{\cellcolor[HTML]{C0C0C0}\textbf{3}} &
  \multicolumn{1}{c|}{\cellcolor[HTML]{C0C0C0}\textbf{4}} \\ \hline
\multicolumn{1}{|c|}{\cellcolor[HTML]{A5E4F7}\textbf{0.05}} &
  \multicolumn{1}{r|}{100.0 \%} &
  \multicolumn{1}{r|}{100.0 \%} &
  95.0 \% &
  \multicolumn{1}{r|}{100.0 \%} &
  \multicolumn{1}{r|}{100.0 \%} &
  100.0 \% &
  \multicolumn{1}{r|}{100.0 \%} &
  \multicolumn{1}{r|}{100.0 \%} &
  100.0 \% \\ \hline
\multicolumn{1}{|c|}{\cellcolor[HTML]{A5E4F7}\textbf{0.10}} &
  \multicolumn{1}{r|}{95.0 \%} &
  \multicolumn{1}{r|}{100.0 \%} &
  95.0 \% &
  \multicolumn{1}{r|}{100.0 \%} &
  \multicolumn{1}{r|}{100.0 \%} &
  100.0 \% &
  \multicolumn{1}{r|}{100.0 \%} &
  \multicolumn{1}{r|}{100.0 \%} &
  100.0 \% \\ \hline
\multicolumn{1}{|c|}{\cellcolor[HTML]{A5E4F7}\textbf{0.25}} &
  \multicolumn{1}{r|}{95.0 \%} &
  \multicolumn{1}{r|}{100.0 \%} &
  80.0 \% &
  \multicolumn{1}{r|}{94.0\%} &
  \multicolumn{1}{r|}{93.0 \%} &
  94.0 \% &
  \multicolumn{1}{r|}{94.8 \%} &
  \multicolumn{1}{r|}{94.0 \%} &
  93.8 \% \\ \hline
\multicolumn{1}{|c|}{\cellcolor[HTML]{A5E4F7}\textbf{0.35}} &
  \multicolumn{1}{r|}{75.0 \%} &
  \multicolumn{1}{r|}{90.0 \%} &
  75.0 \% &
  \multicolumn{1}{r|}{87.0\%} &
  \multicolumn{1}{r|}{88.0 \%} &
  86.0 \% &
  \multicolumn{1}{r|}{88.2 \%} &
  \multicolumn{1}{r|}{88.4 \%} &
  87.2 \% \\ \hline
\end{tabular}
\caption{Percentage of correct evaluations on the test set for PQML model.}
\label{tbl:tab02}
\end{table}

Table \ref{tbl:tab02} shows the percentages of correct evaluations on the test set as a function of the Gaussian Noise on the dataset, the number of qubits used, datasets with different cardinalities (the header of the table shows the number of samples of the test set on the number of total samples). It is referred to PQML model.

\begin{figure}[h!]
    \centering
    \includegraphics[width=1.0\textwidth]{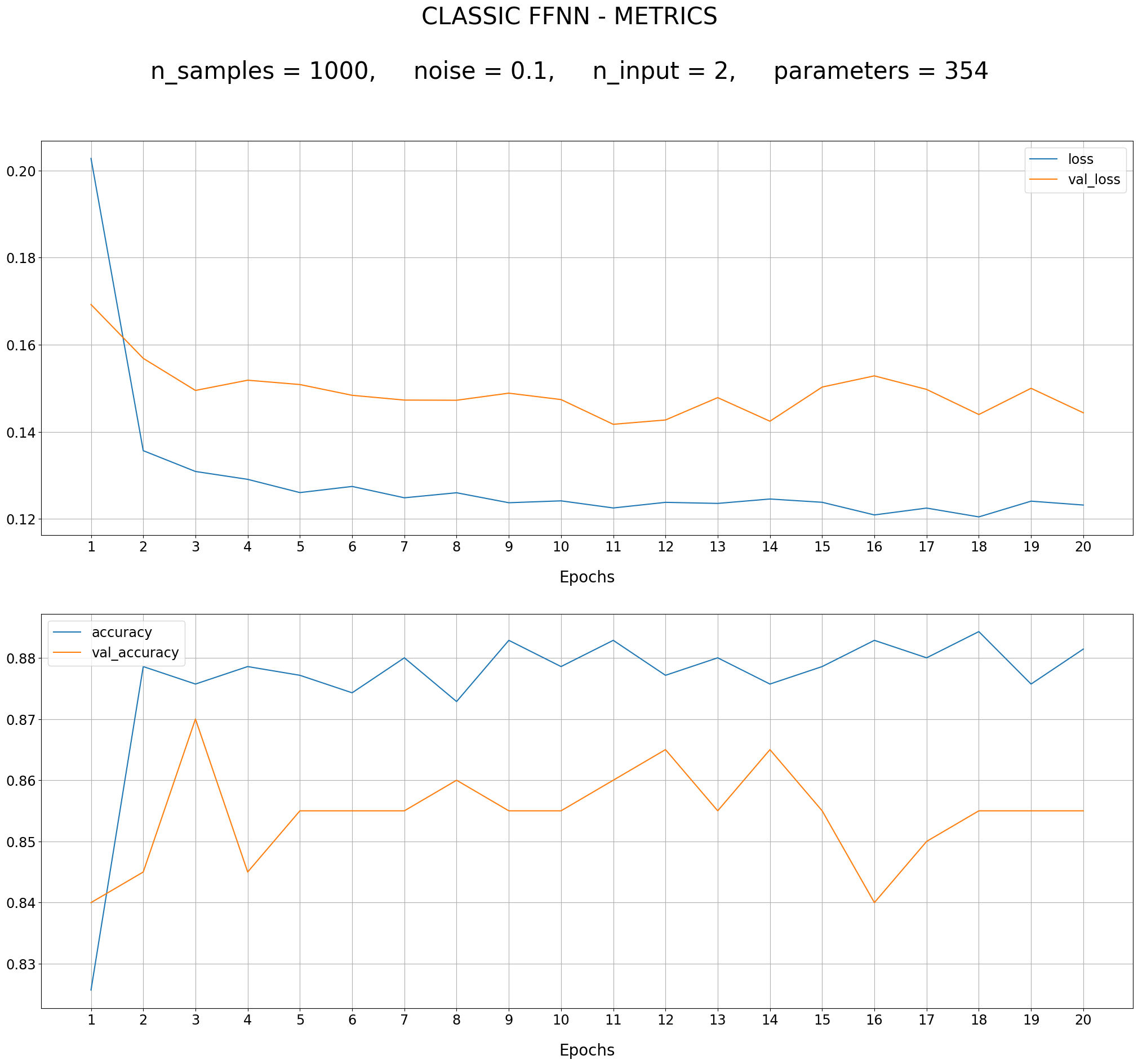}
    \caption{Metrics for an analogue FFNN}
    \label{fig:img07}
\end{figure}

Figure \ref{fig:img07} shows the usual metrics for a simple FFNN Fully-Connected, three dense layers, with the same inputs and output as quantum networks, trained for 20 epochs and with 354 trainable weights.
Without the use of regularisers or drop-out layers, an early overfitting of the model can be observed.

\section{Conclusions and future works}
The two models exhibit remarkably similar tendencies, demonstrating their quality and validity.

Interestingly, unlike an analogous FFNN, a quantum network, since it contains only unitary transformations, can reduce the overfitting phenomenon without the necessity of introducing any particular regularizers or drop-out layers (at least for certain values of ansatz depth).

In our future works, we intend to continue our investigations into quantum neural networks, working on various fronts: new datasets, different ansatz topologies, optimisers and loss functions. Since we focused our attention on some implementation aspects in this paper, ending up working on quantum simulators, we intend to conduct further tests on real machines (e.g. IBM Quantum Experience or Google's Sycamore). We would like to understand better the real effects of quantum noise on circuits.




\printbibliography

\end{document}